\documentstyle[12pt]{article}
\newcommand{\be}{\begin{equation}}
\newcommand{\ee}{\end{equation}}
 
\newcommand{\bea}{\begin{eqnarray}}
\newcommand{\eea}{\end{eqnarray}}
\newcommand{\ba}{\begin{array}}
\newcommand{\ea}{\end{array}}
\newcommand{\beqa}{\begin{eqnarray}}
\newcommand{\eeqa}{\end{eqnarray}}

\newcommand{\NP}[1]{Nucl. Phys.\ {\bf #1}}
\newcommand{\PL}[1]{Phys. Lett.\ {\bf #1}}

\newcommand{\PRD}[1]{Phys. Rev.\ {\bf #1}}
\newcommand{\PR}[1]{Phys. Rep.\ {\bf #1}}
\newcommand{\PRL}[1]{Phys. Rev. Lett.\ {\bf #1}}

\newcommand{\IJMP}[1]{Int. J. Mod. Phys.\ {\bf #1}}

\newcommand{\ZP}[1]{Z. Phys.\ {\bf #1}}

\newcommand{\Tr}{{\rm Tr}}
\renewcommand{\ln}{{\rm ln}}

\newcommand{\sulu}{$SU(2)_L\times U(1)_Y$}

\newcommand{\matr}{\left( \begin{array}}
\newcommand{\ematr}{\end{array} \right)}
 
\newcommand{\lsim}
{{\;\raise0.3ex\hbox{$<$\kern-0.75em\raise-1.1ex\hbox{$\sim$}}\;}}
\newcommand{\gsim}
{{\;\raise0.3ex\hbox{$>$\kern-0.75em\raise-1.1ex\hbox{$\sim$}}\;}} 
 
\begin{document}
 
\begin{titlepage}

\begin{flushright}
{\large HIP-1997-26/TH}\\
{\large hep-ph/9708486}
\end{flushright} 
 
\Large
 
\begin{center}
{\bf Mass of the lightest Higgs Boson in Supersymmetric 
Left-Right Models}
 
\bigskip
\bigskip

\normalsize
{K. Huitu$^a$, P.N. Pandita$^{a,b}$
and K. Puolam\"aki$^a$}\\
[15pt]
{\it $^a$Helsinki Institute of Physics, P.O.Box 9, 
FIN-00014 University of Helsinki,
Finland
\\$^b$Department of Physics, North Eastern Hill University,
Shillong 793022, India\footnote{Permanent address} }

\vspace{1.5truecm}

{\bf\normalsize \bf Abstract}

 \end{center}

{\normalsize
We consider the lightest Higgs boson in naturally R-parity conserving
supersymmetric left-right models.
We obtain an upper bound on the tree level mass of 
this lightest Higgs boson. This upper 
bound depends on the $SU(2)_L$ and $SU(2)_R$ gauge couplings,
and the vacuum expectation values of bidoublet Higgs fields, which
are needed to break $SU(2)_L\times U(1)_Y$.
The upper bound does not depend on either the $SU(2)_R$ breaking scale
or the supersymmetry breaking scale.
We evaluate the bound numerically by assuming that the theory 
remains perturbative upto some scale $\Lambda$.
We find that the bound can be considerably larger than in MSSM. 
The dominant radiative corrections to the upper bound due to top-stop
and bottom-sbottom systems are of the same form as in the
minimal supersymmetric standard model.

}

\vfill

\normalsize
\noindent

\end{titlepage}
 
\newpage
 
\setcounter{page}{2}

Considerable importance attaches to the study of the Higgs bosons of the
minimal supersymmetric standard model (MSSM), based on the Standard Model
gauge group $SU(2)_L\times U(1)_Y$, with two Higgs doublet superfields
\cite{susy}.
It is well known that, because of underlying gauge 
invariance and supersymmetry
(SUSY), the lightest Higgs boson of MSSM has a tree 
level upper bound of $m_Z$
(the mass of Z boson) on its mass \cite{mass1}.
Although radiative corrections \cite{rad} to the tree 
level result can be
appreciable, these depend only logarithmically on the SUSY 
breaking scale, 
and are, therefore, under control.
This results in an upper bound of about 125
- 135 GeV on the  radiatively corrected mass of the lightest 
Higgs boson of MSSM \cite{mass2}.
Because of the presence of additional trilinear Yukawa couplings, 
such a tight
constraint on the mass of the lightest Higgs boson need 
not {\it a priori}  hold in
extensions of MSSM with an extended Higgs sector.
Nevertheless, it has been shown that the upper bound 
on the lightest Higgs boson mass  in these models depends only  
on the weak scale and
dimensionless coupling constants (and only 
logarithmically on SUSY breaking
scale), and is calculable if all the couplings remain perturbative
below some scale $\Lambda$ [5-11].
This upper bound can vary between 150 GeV to 165 GeV depending 
on the Higgs
structure of the supersymmetric model.
Thus, nonobservation of a light Higgs boson below this 
upper bound will rule
out an entire class of supersymmetric models based 
on the gauge group \sulu .

The existence of the upper bound on the lightest Higgs 
boson mass in MSSM with
arbitrary Higgs sectors has been investigated in a 
situation where the underlying supersymmetric model respects 
the discrete matter parity, or
R-parity ($R_P$) symmetry \cite{r1,r2}, under which all 
Standard Model particles are even and their superpartners are odd, 
so that all renormalizable
baryon (B) and lepton (L) number violating terms in the 
superpotential are forbidden.
However, the assumption of R-parity conservation 
appears to be {\it ad hoc},
since it is not required for the internal consistency of 
the minimal supersymmetric standard model.
Furthermore, all global symmetries, discrete or continuous, 
could be violated by the Planck scale physics effects \cite{planck}.
The problem becomes accute for low energy supersymmetric models
\cite{Rpar}, because B and L are no longer automatic 
symmetries of the Lagrangian as they are in the Standard Model.

It would, therefore, be more appealing to have a supersymmetric 
theory where R-parity is related to a gauge symmetry, and its 
conservation is automatic because of the invariance of the 
underlying theory under an extended gauge symmetry.
Indeed $R_P$ conservation follows automatically in certain theories 
with gauged $(B-L)$, as is suggested by the fact that 
matter parity is simply a $Z_2$ subgroup of $(B-L)$.
It has been noted by several authors \cite{Rauto1,Rauto2} 
that if the gauge symmetry of MSSM is 
extended to $SU(2)_L\times U(1)_{I_{3R}}\times U(1)_{B-L}$,
or $SU(2)_L\times SU(2)_R\times U(1)_{B-L}$, the theory 
becomes automatically R-parity conserving.
Such a supersymmetric left-right theory (SUSYLR) 
solves the problems of explicit B and L 
violation of MSSM, and has received much attention recently
[18-23].

Since such a naturally R-parity conserving theory 
necessarily involves the extension of the 
Standard Model gauge group, and since the extended gauge
symmetry has to be broken, it involves a {\it new scale}, 
the scale of left-right symmetry breaking, beyond the SUSY 
and $SU(2)_L\times U(1)_Y$ breaking scales of MSSM.
It is, therefore, important to ask whether the upper bound 
on the lightest Higgs mass in naturally R-parity conserving 
theories depends on the scale of
the breakdown of the extended gauge group.
In this paper we investigate the Higgs sector of the 
supersymmetric left-right theory in order to answer this question.
We find that the tree level upper bound on the lightest Higgs boson
mass does not explicitly depend on the scale of the left-right 
symmetry breaking.
It depends on the scale of the $SU(2)_L\times U(1)_Y$ breaking
and dimensionless coupling constants only.
We then calculate the dominant one-loop radiative corrections 
due to top-stop and bottom-sbottom to this upper bound on the 
lightest Higgs mass in the supersymmetric left-right models.
These turn out to be of the same order of magnitude as 
the corresponding radiative corrections in models based 
on $SU(2)_L\times U(1)_Y$.
Although the R-parity is conserved at the level of Lagrangian, it
is necessarily spontaneously broken in this class of models 
\cite{km1}.
We will see that the upper bound does not depend on the VEV of the
right-handed sneutrino responsible for the spontaneous
R-parity violation.

We begin by recalling the basic features of the left-right 
supersymmetric models.
The quark and lepton doublets are included in
$Q(2,1,1/3)$; $Q^c(1,2,-1/3)$; $L(2,1,-1)$; $L^c(1,2,1)$, 
where $Q$ and $Q^c$
denote the left- and right-handed quark superfields 
and similarly for the leptons $L$ and $L^c$.
The Higgs superfields consist of 
$\Delta_L(3,1,-2)$; $\Delta_R(1,3,-2)$; 
$\delta_L(3,1,2)$; $\delta_R(1,3,2)$;  
$\Phi(2,2,0)$; $\chi(2,2,0)$.
The numbers in the parentheses denote the representation 
content of the fields under the gauge group 
$SU(2)_L\times SU(2)_R\times U(1)_{B-L}$.
We note that two $SU(2)_R$ Higgs triplet superfields 
$\Delta_R (1,3,-2)$ and $\delta_R(1,3,2)$ with 
opposite $(B-L)$ are necessary to break the
left-right symmetry spontaneously, and to cancel 
triangle gauge anomalies due to the fermionic superpartners.
The left-right model also contains the $SU(2)_L$ triplets 
$\Delta_L$ and $\delta_L$ in order to make the Lagrangian fully 
symmetric under the  $L\leftrightarrow R$ transformation, 
although these are not needed phenomenologically for 
the symmetry breaking or the see-saw mechanism.

We further note that there are two bidoublet Higgs 
superfields $\Phi$ and $\chi$ in order to 
break the $SU(2)_L\times U(1)_Y$ 
and to generate a nontrivial Kobayashi-Maskawa matrix.
The most general gauge invariant superpotential involving 
these superfields can be written as
\bea
W&=& h_{\phi Q}Q^T i\tau_2 \Phi Q^c +
h_{\chi Q}Q^T i\tau_2 \chi Q^c +
h_{\phi L}L^T i\tau_2 \Phi L^c 
+ h_{\chi L}L^T i\tau_2 \chi L^c \nonumber\\
&&+h_{\delta_L} L^T i\tau_2 \delta_L L +
h_{\Delta_R} L^{cT} i\tau_2 \Delta_R L^c+
\mu_1 \Tr (i\tau_2\Phi^T i\tau_2 \chi) +
\mu_1' \Tr (i\tau_2\Phi^T i\tau_2 \Phi) \nonumber\\
&&+ \mu_1'' \Tr (i\tau_2\chi^T i\tau_2 \chi) 
+\Tr (\mu_{2L}\Delta_L \delta_L +
\mu_{2R}\Delta_R\delta_R).
\label{superpot}
\eea
The general form of the Higgs potential is given by
\be
V=V_F+V_D+V_{soft}
\label{pot}
\ee
\noindent
which can be calculated in a straightforward manner.
In the following we shall represent the scalar components 
of the Higgs superfields
by the same symbols as the superfields themselves and add
a tilde on the scalar components of lepton and quark superfields.
The most general form of the vacuum expectation values 
of various scalar fields, which preserves $U(1)_{em}$ 
can be written as

\bea
&&\langle \Phi\rangle  = \matr {cc} \kappa_1&0\\
0&e^{i\varphi_1}\kappa '_1 \ematr ,\;\;
\langle \chi\rangle  = \matr {cc} e^{i\varphi_2}\kappa '_2&0\\
0&\kappa_2 \ematr ,
\nonumber\\
&& \langle \Delta_L\rangle = \matr {cc} 0&v_{\Delta_L}\\
0&0\ematr ,\;
\langle \delta_L\rangle = \matr {cc} 0&0\\v_{\delta_L}&0\ematr , 
\nonumber\\
&&\; \langle \Delta_R\rangle = \matr {cc} 0&v_{\Delta_R}\\
0&0\ematr ,\;
\langle \delta_R\rangle = \matr {cc} 0&0\\v_{\delta_R}&0\ematr ,
\nonumber\\
&&\langle L\rangle =\matr {c} \sigma_L\\0\ematr,\;
\langle L^c\rangle =\matr {c} 0\\\sigma_R\ematr .
\eea

We note that the triplet vacuum expectation values 
$v_{\Delta_R}$ and $v_{\delta_R}$ represent 
the scale of $SU(2)_R$ breaking and are, 
therefore, assumed to be large, in the range $v_{\Delta_R},\; 
v_{\delta_R}\gsim 1$ TeV.
These represent a new scale of physics, the right-handed scale 
which we shall generically denote as $v_R$.
Since the mixing between the charged gauge bosons is tiny, 
and to avoid the flavor changing neutral currents, 
$\kappa'_1$ and $\kappa'_2$ are taken to be much smaller than 
$\kappa_1$ and $\kappa_2$, and
we shall ignore them in the following.
Furthermore, since the electroweak $\rho$-parameter 
is close to unity, $\rho=1.002\pm 0.0013\pm 0.0018$ \cite{pdg}, 
the triplet vacuum expectation values $\langle \Delta_L\rangle $ and 
$\langle\delta_L\rangle $ must be small, and we shall ignore 
them as well. Since the spontaneous 
breakdown of R-parity is inevitable \cite{km1}, we shall assume 
that at least one of the VEVs $\langle \tilde\nu\rangle$ or 
$\langle \tilde\nu^c \rangle$ is non-zero.
The VEV $\langle \tilde\nu \rangle\equiv\sigma_L$ is at most 
of the order of the weak
scale, since it contributes to lighter weak gauge boson masses.
It is important to note that in $Q_{em}$ preserving ground state
$\langle \tilde\nu^c \rangle \equiv\sigma_R$ is necessarily at 
least of the order of the typical SUSY breaking scale $M_{SUSY}$ 
or the right-handed breaking scale ($v_{\Delta_R},\; v_{\delta_R}$), 
whichever is lower \cite{km3}.

We now proceed to the main point of this paper by constructing the 
mass matrix for the neutral scalars.
To this end we write down explicitly the different components of the 
scalar
potential (\ref{pot}) ($g_L,\; g_R,\;g_{B-L}$ are the gauge 
couplings)

\bea
V_F&=&
+|h_{\phi L} i\tau_2 \phi L^c +h_{\chi L} i\tau_2 \chi L^c
+2h_{\delta_L}  L^{T}i\tau_2 \delta_L|^2 \nonumber\\
&&
+|h_{\phi L} L^T i\tau_2\phi + h_{\chi L} L^T i\tau_2\chi
+2h_{\Delta_R}  L^{cT}i\tau_2 \Delta_R |^2 \nonumber \\
&&
+|h_{\Delta_R} L^c  L^{cT} (i\tau_2) +\mu_{2R} \delta_R |^2
+|h_{\delta_L} L  L^{T} (i\tau_2) +\mu_{2L} \Delta_L |^2
\nonumber\\
&&
+|h_{\phi Q}Q^cQ^T(i\tau_2)+h_{\phi L}L^cL^T(i\tau_2)+
\mu_1(i\tau_2)\chi^T(i\tau_2) +2\mu'_1(i\tau_2)\Phi^T(i\tau_2) |^2
\nonumber\\
&&+
|h_{\chi Q}Q^cQ^T(i\tau_2)+h_{\chi L}L^cL^T(i\tau_2)+
\mu_1(i\tau_2)\Phi^T(i\tau_2) + 2\mu''_1(i\tau_2)\chi^T(i\tau_2)|^2
\nonumber\\
&& + |(i\tau_2)(h_{\phi Q}\Phi +h_{\chi Q}\chi )Q^c|^2+
|Q^T(i\tau_2)(h_{\phi Q}\Phi +h_{\chi Q}\chi )|^2\nonumber\\
&&+|\mu_{2R}\Delta_R|^2 +|\mu_{2L}\delta_L|^2,
\label{Fterm}
\eea

\bea
V_D&=& \frac 18 g_L^2\sum_a \left[ \Tr (\Phi^\dagger\tau_a\Phi )+
\Tr (\chi^\dagger\tau_a\chi )
+2\Tr (\Delta_L^\dagger\tau_a\Delta_L )
+2\Tr (\delta_L^\dagger\tau_a\delta_L )\right.
\nonumber\\
&&\left. +L^\dagger\tau_a L+Q^\dagger\tau_a Q \right]^2
+\frac 18 g_R^2\sum_a \left[ -\Tr (\Phi\tau_a\Phi^\dagger )-
\Tr (\chi\tau_a\chi^\dagger )\right.  \nonumber\\
&&\left. +2\Tr (\Delta_R^\dagger\tau_a\Delta_R )
+2\Tr (\delta_R^\dagger\tau_a\delta_R )
+L^{c\dagger}\tau_a L^c+Q^{c\dagger}\tau_a Q^c \right]^2
\nonumber\\
&&+\frac 18 g_{B-L}^2\left[ 2\Tr (-\Delta_R^\dagger\Delta_R+
\delta_R^\dagger\delta_R-\Delta_L^\dagger\Delta_L+
\delta_L^\dagger\delta_L)\right.
\nonumber\\ &&
\left. -L^\dagger L+L^{c\dagger} L^c
+\frac 13 Q^\dagger Q-\frac 13 Q^{c\dagger} Q^c \right]^2,
\label{Dterm}
\eea

\noindent
and

\bea
V_{soft}&=&
m_1^2\Tr |\phi|^2 + m_2^2\Tr |\chi|^2 -
(m_{\phi \chi}^2  \Tr (i\tau_2 \phi^T i\tau_2 \chi )
+m_{\phi \phi}^2  \Tr (i\tau_2 \phi^T i\tau_2 \phi )\nonumber\\
&&+m_{\chi \chi}^2  \Tr (i\tau_2 \chi^T i\tau_2 \chi ) +h.c.)
+m_{3 }^2 |\Delta_R |^2 + m_{4 }^2|\delta_R |^2
-(m_{\Delta\delta}^2\Tr \Delta_R\delta_R +h.c.)\nonumber\\
&&+m_{5 }^2 |\Delta_L |^2 + m_{6 }^2|\delta_L |^2
-({m_{\Delta\delta}}^{'2}\Tr \Delta_L\delta_L +h.c.)
+m_7^2 | L^c|^2
+m_8^2 |L|^2 \nonumber\\
&&+ (L^{T}i\tau_2 (A_\phi\phi +A_\chi\chi ) L^c
+A_{\Delta_R} L^{cT} i\tau_2 \Delta_R L^c
+A_{\delta_L} L^{T} i\tau_2 \delta_L L +h.c. )
\nonumber\\
&&+m_9^2|Q|^2+ m_{10}^2|Q^c|^2+
(Q^Ti\tau_2(B_\phi \Phi +B_\chi \chi )Q^c +h.c.).
\label{softterm}
\eea

\noindent
{}From (\ref{pot}), (\ref{Fterm}), (\ref{Dterm}) and (\ref{softterm})
it is straightforward to
derive the mass matrix for the CP-even neutral scalars, whose 
eigenvalues will provide the masses of the physical scalar 
Higgs particles. We shall not write the full $10\times 10$ mass 
matrix \cite{hpp2}  here,
since for the specific purpose of the determination 
of a general bound on the
lightest Higgs mass, the problem is much simpler.
A known property of any Hermitian matrix is that its minimum 
eigenvalue must
be smaller than that of its upper left corner $2\times 2$ 
submatrix. Using this fact, and calling $m_{ij}$ the matrix 
elements of the CP-even
neutral Higgs mass squared matrix, we can write the following
rigorous bound on the squared mass ($m_{h}^2$) of the lightest 
Higgs boson:

\be
m_{h}^2\leq \frac{m_{11}^2+m_{22}^2} 2
\left[1-\sqrt{1-4\frac{m_{11}^2m_{22}^2 - m_{12}^4}
{m_{11}^2+m_{22}^2}}\; \right].
\label{hmgen}
\ee

We shall choose a basis for the Higgs mass matrix such that the 
first two indices correspond to ($\Phi_1^0,\chi^0_2$).
Starting from the expression (\ref{pot})-(\ref{softterm}) for 
the potential and imposing the minimization conditions 
$\partial V/\partial \Phi^0_1 = \partial V/\partial \chi^0_2 = 0$, 
we obtain the following expression for
$m_{11}^2,\; m_{22}^2,\; m_{12}^2$ 
(from now on we will assume that only one R-parity violating
vacuum expectation value, namely $\sigma_R\neq 0$):

\bea
m_{11}^2&=& -m_{\Phi\chi }^2\frac{\kappa_2}{\kappa_1}
+\frac 12 (g_L^2+g_R^2)\kappa_1^2,\nonumber\\
m_{22}^2&=& -m_{\Phi\chi }^2\frac{\kappa_1}{\kappa_2}
+\frac 12 (g_L^2+g_R^2)\kappa_2^2,\nonumber\\
m_{12}^2&=& m_{\Phi\chi }^2
-\frac 12 (g_L^2+g_R^2)\kappa_1\kappa_2.
\label{hmm}
\eea

\noindent
{}From eqs. (\ref{hmgen})-(\ref{hmm}), we immediately obtain 
the upper bound on the lightest Higgs boson mass in the left-right 
supersymmetric model:

\be
m_{h}^2\leq \frac 12 (g_L^2+g_R^2)(\kappa_1^2 +\kappa_2^2)\cos^2 
2\beta =
\left( 1+\frac{g_R^2}{g_L^2}\right) m_{W_L}^2 \cos^2 2\beta,
\label{hm}
\ee

\noindent
where $\tan\beta = \kappa_2/\kappa_1 $. 
We note that the upper bound (\ref{hm}) is 
independent both of the supersymmetry breaking parameters 
(as in the case of supersymmetric models based on 
$SU(2)_L\times U(1)_Y$) and of the
$SU(2)_R$ breaking scale, which, {\it a priori} can be very large.
It is also independent of the $R$-parity breaking vacuum expectation  
value $\sigma_R$. The upper bound is  controlled by the weak scale
and the
dimensionless gauge couplings $(g_L$ and $g_R$).\footnote{
Even if we do not neglect $\kappa_1',\;\kappa_2'$ and $\sigma_L$,
the upper bound  on the lightest Higgs mass does not depend either 
on the supersymmetry breaking scale, or 
right-handed breaking scale, or $\sigma_R$ \cite{hpp2}, 
although it depends on $\sigma_L$. However, 
since $\sigma_L$ contributes to the lighter $W$ mass, it is  
at most of the order of weak scale.}
Since the right-handed gauge coupling $g_R$ is not known, 
the upper  bound
on the right-hand side of (\ref{hm}) comes from the 
requirement that the
left-right supersymmetric model remains perturbative 
below some scale $\Lambda $.
In order to implement this requirement we need to solve the 
renormalization
group equations (RGE's) for the gauge couplings of the theory.
We do not require that the gauge couplings unify at 
some scale.

\input{epsf.sty}
\begin{figure}[t]
\leavevmode
\begin{center}
\mbox{\epsfxsize=10.cm\epsfysize=10.cm\epsffile{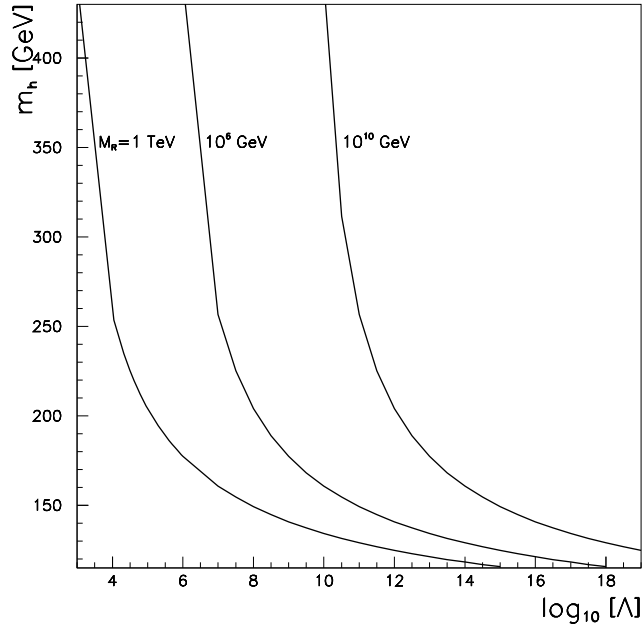}}
\end{center}
\caption{The tree level upper bound on the lightest Higgs mass as
a function of the scale $\Lambda$ upto which the $g_R$ coupling 
remains perturbative.
The plotted $SU(2)_R\times U(1)_{B-L}$ breaking scales are
$M_R=1$ TeV, $10^6$ GeV and $10^{10}$ GeV.
}
\end{figure}

In the left-right supersymmetric theory defined by the superpotential
(\ref{superpot}), the one-loop RGE's for the
$SU(2)_L\times SU(2)_R\times U(1)_{B-L}$ gauge couplings 
can be written as
\bea
16\pi^2 \frac{dg_{L,R}}{dt}&=&6g_{L,R}^3 , \nonumber\\
16\pi^2 \frac{dg_{B-L}}{dt}&=&16g_{B-L}^3.
\eea

\noindent
Requiring that $g_R$ remains perturbative upto a scale $\Lambda$
($g_i^2(Q^2)/4\pi\leq 1$ for $Q^2\leq \Lambda^2$, where the equality
holds for $Q^2 =\Lambda^2$), we
obtain the upper bound on the lightest Higgs mass as shown in
Fig. 1 for different values of $\Lambda $ ranging from $\Lambda =1$
TeV to $\Lambda = 10^{19}$ GeV and $SU(2)_R\times U(1)_{B-L}$ 
breaking
scales $M_R=1$ TeV, $10^6$ GeV and $10^{10}$ GeV, respectively.
The gauge couplings $g_L$, $g_R$ and $g_{B-L}$ are connected
by the 
weak mixing angle, $\tan\theta_W =
{g'}/{g_L}=g_Rg_{B-L}/g_L\sqrt{g_R^2+g_{B-L}^2}$.
We have checked that for the values used, also $g_{B-L}$ remains
perturbative upto the scale $\Lambda$.
The coupling $g_R$ has a lower limit from
$\sin^2\theta_W=e^2/g_L^2\simeq 0.23$, namely
$g_R\geq 0.55 g_L$ \cite{cl}, which is also fulfilled in Fig.1.
Due to the lower limit on $g_R$, the smallest upper bound
is given by $m_{h}\lsim 92$ GeV $\times |\cos 2\beta|$.
Since the bound (\ref{hm}) increases with increasing $g_R$, 
it becomes less
restrictive for smaller values of $\Lambda$ or larger 
values of $M_R$. If the difference between $\Lambda$ and $M_R$ 
is larger than two orders of
magnitude, the tree-level upper bound remains 
below $\sim 205$ GeV.
When $\Lambda =M_R$ and hence $g_R^2(\Lambda^2) =4\pi$, the bound
is at its largest, namely $m_{h}\lsim 446$ GeV.

We now proceed to the calculation of dominant one-loop 
radiative corrections to the upper bound (\ref{hm}) 
on the lightest Higgs boson mass in SUSYLR model.
We shall use the method of one-loop effective potential 
\cite{cw} for the calculation of radiative corrections, 
where the effective potential may be expressed as the sum 
of tree-level potential plus a correction coming 
from the sum of one-loop diagrams with external 
lines having zero momenta,

\be
V_{1-loop}=V_{tree} +\Delta V_1,
\ee

\noindent
where $V_{tree}$ is the tree level potential (\ref{pot}) evaluated at
the appropriate running scale $Q$, and $\Delta V_1$ is the one-loop 
correction given by

\be
\Delta V_1
=\frac 1{64\pi^2}\sum_i (-1)^{2J_i} (2J_i+1)
m_i^4 (\ln \frac {m_i^2}{Q^2} -\frac 32 ).
\label{v1}
\ee

\noindent
In (\ref{v1}), $m_i$ is the mass of the $i$th particle with spin 
$J_i$ in the appropriate background.
The dominant contribution to (\ref{v1}) arises from the top-stop
($t -\tilde t $) and bottom-sbottom ($b-\tilde b $) systems.
For the degenerate stop case the one-loop correction is given by

\bea
\Delta V_1
&=&\frac 3{16\pi^2}\left[(\tilde m^2 + h_t^2|\kappa_2|^2)^2 
\left(
\ln \frac{\tilde m^2 + h_t^2|\kappa_2|^2}{Q^2} - 
\frac 32\right)\right.
\nonumber\\
&&
\left.
-h_t^4|\kappa_2|^4 \left(
\ln \frac{h_t^2|\kappa_2|^2}{Q^2} - \frac 32\right) \right],
\label{vv1}
\eea

\noindent
with an analogous expression for the sbottom case.
In (\ref{vv1}) $\tilde m^2$ is the SUSY breaking mass for the 
squarks, $Q$ is the renormalization scale, and $h_t$ is the 
top Yukawa coupling ($h_t=(h_{\chi Q})_{33}$, see 
(\ref{superpot})). In the case with a large left-right squark 
mixing the formula (\ref{vv1}) should be extended appropriately 
\cite{hpp2}. In numerical calculations, we shall use the full 
formula which includes the left-right mixing in stop and 
sbottom mass matrices.

\begin{figure}[b]
\leavevmode
\begin{center}
\mbox{\epsfxsize=15.truecm\epsfysize=15.truecm\epsffile{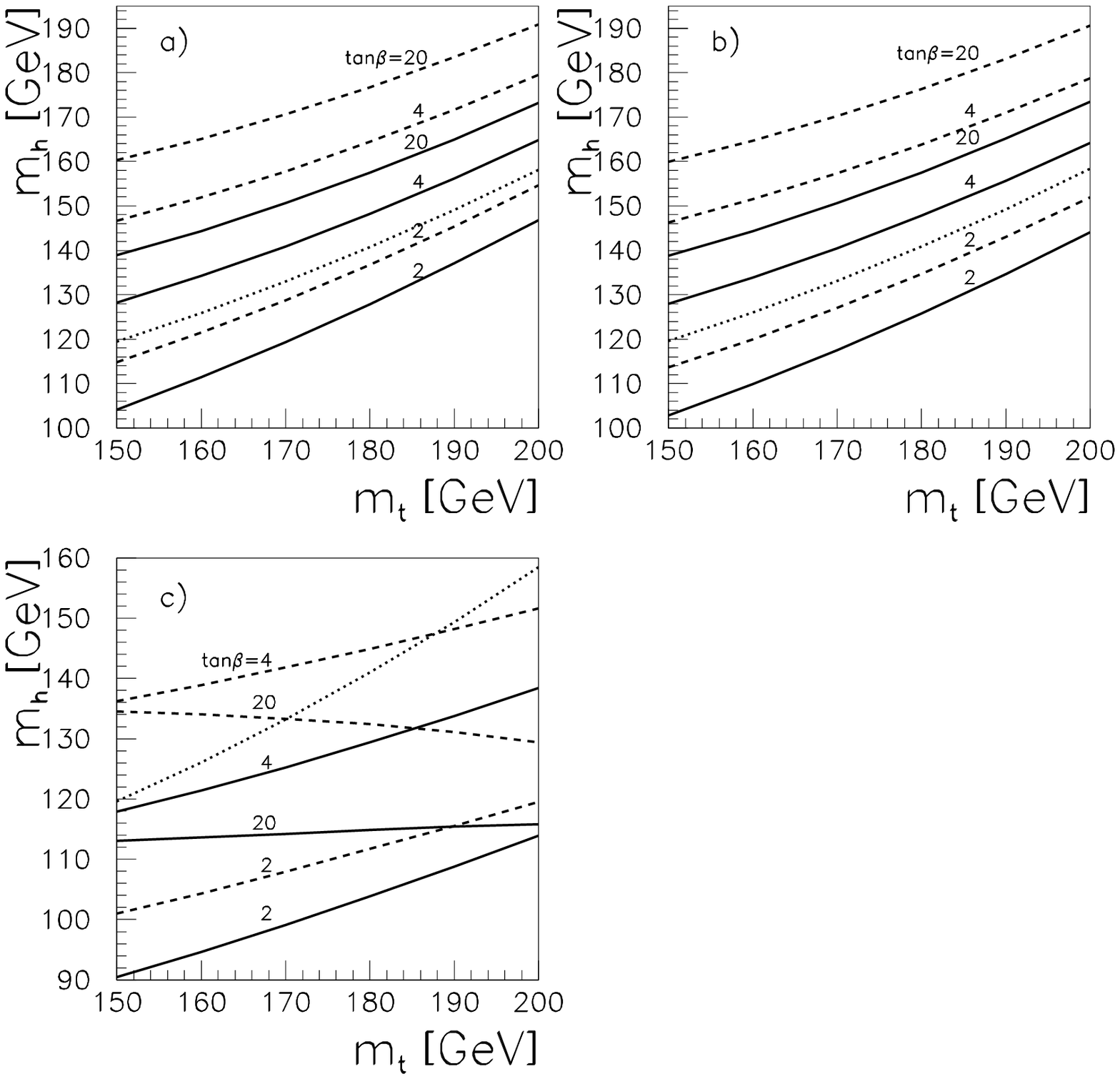}}
\end{center}
\caption{ The radiatively corrected upper limit on the mass of
the lightest Higgs boson as a
function of $m_t$ with $\Lambda = 10^{16}$ GeV and $A_t=A_b=1$ =
TeV. 
The solid line corresponds to the $SU(2)_R$ scale
of 10 TeV and the dashed line to the  $SU(2)_R$ scale
of $10^{10}$ GeV.
The dotted curve corresponds to MSSM limit for $\tan\beta =20$
and $\mu=\mu_1$. 
In a)  $\mu_1=\mu'_1=\mu''_1=0$,
in b)  $\mu_1=\mu'_1=\mu''_1=500$ GeV, and
in c) $\mu_1=\mu'_1=\mu''_1=1000$ GeV.
}
\end{figure}

Using (\ref{vv1}), one can derive the radiatively corrected upper 
bound on the mass of the lightest CP-even Higgs boson in 
the left-right supersymmetric model:

\be
m_{h}^2\leq \frac 12 \left[ (g_L^2+g_R^2)(\kappa_1^2+\kappa_2^2)
\cos^2 2\beta +
G(\Delta_{11}\cos^2\beta +\Delta_{22}\sin^2\beta +\Delta_{12}\cos 
2\beta )
\right],
\label{mhb}
\ee

\noindent
where $G=3g_L^2/(8\pi^2m_W^2)$, and $\Delta_{ij}$ ($i,j=1,2$), 
which signify radiative corrections, are complicated functions 
of the parameters of
the model \cite{hpp2}, and are similar in structure to the 
corresponding
quantities in the models based on $SU(2)_L\times U(1)_Y$ gauge
group \cite{erz}.
For the degenerate stop and sbottom case, they reduce to
\bea
\Delta_{11}=\frac{m_b^4}{\cos^2\beta}
\ln\left(\frac{m^2_{\tilde b_1}m^2_{\tilde b_2}}{m_b^4}\right),
\;\; \Delta_{22}=\frac{m_t^4}{\sin^2\beta}
\ln\left(\frac{m^2_{\tilde t_1}m^2_{\tilde t_2}}{m_t^4}\right),
\eea
with $\Delta_{12}=0$, i.e. the same as in MSSM.
In Fig. 2, we have plotted the upper bound (\ref{mhb}) 
as a function of
top quark mass in the range $150<m_t<200$ GeV, which subsumes the 
recent
direct measurement (CDF and D0 combined) from the Tevatron
$p\bar p$ collider of $m_t=175\pm 6$ GeV \cite{grannis}.

In Fig. 2 we have taken the soft supersymmetry breaking trilinear 
couplings to be $A_t=A_b=1$ TeV, where
$A_t=(B_{\Phi} h_{\Phi Q}^{-1})_{33}$, $A_b=(B_{\chi} h_{\chi Q}^{-1})_{33}$.
The solid line corresponds to $SU(2)_R\times U(1)_{B-L}$ breaking
scale 10 TeV and dashed one to $10^{10}$ GeV.
The bound on the lightest Higgs mass becomes less restrictive
for larger $SU(2)_R$ breaking scales.
In Fig. 2 a) we have taken $\mu_1=\mu'_1=\mu''_1=0$.
The MSSM limit with $\mu=0$ and $\tan\beta = 20$ is plotted
as dotted line.
In all cases $\Lambda = 10^{16}$ GeV.
The upper bound increases with increasing $M_R$ scale.
For $M_R=10$ TeV and $m_{top}=175$ GeV, the bound remains below 
155 GeV while for $M_R=10^{10}$ GeV it remains below 175 GeV.
In Figures 2 b) and c) the dependence on $\mu_1,\mu'_1$ and $\mu''_1$
is shown.
The dependence on $\mu $-type parameters 
is enhanced compared to the MSSM expressions
\cite{erz}, since $-\mu\tan\beta$ is replaced by
$(\mu_1+2\mu''_1\frac{m_t}{m_b}\cot\beta )\tan\beta $.
Except for large $\mu_1,\mu'_1,\mu''_1$, it is seen that the mass limits
are somewhat higher in SUSYLR than in the MSSM.
Compared to the limits from models with gauge singlets
\cite{pp,kot,ec}, one finds that, depending on the 
unknown couplings, the
limit may be smaller or larger.

To conclude we have obtained an upper bound on the lightest 
Higgs boson  mass in the supersymmetric left-right model, and have 
shown that it does not depend on the soft supersymmetry breaking 
parameters or potentially large $SU(2)_R\times U(1)_{B-L}$ breaking 
scales.
Furthermore it does not depend on the vacuum expectation value
of the right-handed sneutrino.
The tree-level bound, however, can be considerably larger 
than the corresponding bound in MSSM, if the difference between 
the high scale $\Lambda$ 
and the intermediate scale $M_R$ is small.
The radiative corrections to the upper bound 
from top-stop and bottom-sbottom 
sector are sizable and of the same form as in the MSSM.

\vspace{2.truecm}
\noindent
{\bf Acknowledgements}

One of us (PNP) would like to thank the Helsinki Institute of 
Physics for
hospitality while this work was completed.
The work of PNP is supported by the Department of Atomic Energy 
Project No.37/14/95-R \& D-II/663.

\end{document}